\documentclass[showpacs,10pt,twocolumn,prl]{revtex4-1}
\usepackage{amsmath}
\usepackage{amssymb}
\usepackage{graphics}
\usepackage{epsfig}
\usepackage{color}

\begin{document}
	
	
\title{Quantum transport evidence of topological band structures of kagome superconductor CsV$_{3}$Sb$_{5}$}
\author{Yang Fu$^{1,\dag}$, Ningning Zhao$^{1,\dag}$, Zheng Chen$^{2,\dag}$, Qiangwei Yin$^{1,\dag}$, Zhijun Tu$^{1}$, Chunsheng Gong$^{1}$, Chuanying Xi$^{2}$, Xiangde Zhu$^{2,*}$, Yuping Sun$^{2,3,4}$, Kai Liu$^{1,*}$, and Hechang Lei$^{1,*}$}
\affiliation{$^{1}$Department of Physics and Beijing Key Laboratory of Opto-electronic Functional Materials $\&$ Micro-nano Devices, Renmin University of China, Beijing 100872, China\\
$^{2}$Anhui Province Key Laboratory of Condensed Matter Physics at Extreme Conditions, High Magnetic Field Laboratory, HFIPS, Chinese Academy of Sciences, Hefei 230031, China\\
$^{3}$Key Laboratory of Materials Physics, Institute of Solid State Physics, HFIPS, Chinese Academy of Sciences, Hefei 230031, China\\
$^{4}$Collaborative Innovation Centre of Advanced Microstructures, Nanjing University, Nanjing 210093, China
}
	
\date{\today}
	
\begin{abstract}
We report the transport properties of kagome superconductor CsV$_{3}$Sb$_{5}$ single crystals at magnetic field up to 32 T. The Shubnikov de Haas (SdH) oscillations emerge at low temperature and four frequencies of $F_{\alpha}=$ 27 T, $F_{\beta}=$ 73 T, $F_{\epsilon}=$ 727 T, and $F_{\eta}=$ 786 T with relatively small cyclotron masses are observed. For $F_{\beta}$ and  $F_{\epsilon}$, the Berry phases are close to $\pi$, providing a clear evidence of nontrivial topological band structures of CsV$_{3}$Sb$_{5}$. 
Furthermore, the consistence between theoretical calculations and experimental results implies that these frequencies can be assigned to the Fermi surfaces locating near the boundary of Brillouin zone and confirms that the structure with an inverse Star of David distortion could be the most stable structure at charge density wave state.
These results will shed light on the nature of correlated topological physics in kagome material CsV$_{3}$Sb$_{5}$.
\end{abstract}
	

\maketitle
	
Because of special lattice geometry and multiple sublattices in a unit cell,  materials with a kagome lattice exhibit many of novel physical properties. For example, insulating magnetic kagome materials with strongly geometrical frustration are very promising systems to realize quantum spin liquid state with fractionalized excitations \cite{Balents}.
In contrast, kagome metals exhibit nontrivial topological electronic structures, like Dirac or Weyl nodal points and flat bands \cite{YeL,LiuZ,WangQ2}. 
When combined with long-range magnetism, many of exotic phenomena appear in the magnetic kagome metals, such as large anomalous Hall effect (AHE) \cite{Nakatsuji,WangQ,YeL,LiuE,WangQ2}, negative magnetism of flat band \cite{YinJX}, and large magnetic-field tunability \cite{YinJX2,LiY}.
Thus, kagome metals have become an important platform to study novel physics of correlated topological materials.
	
Recently, the coexistence of charge density wave (CDW) state and superconductivity has been discovered in $A$V$_{3}$Sb$_{5}$ ($A$ = K, Rb and Cs) with the kagome lattice of V atoms, which also have a nonzero $Z_{2}$ topological invariant \cite{Ortiz1,Ortiz2,Ortiz3,YinQW}. 
Further studies indicate that there is a three-dimensional (3D) 2$\times$2$\times$2 superlattice at CDW state \cite{LiangZ,LiHX}, which could lead to an inverse Star of David (ISD) distortion in kagome lattice and a possible chiral charge order accompanying with large anomalous Hall conductivity even the long-range magnetic order is absent \cite{JiangYX,TanH,FengX,Denner,YangSY,YuFH,Kenney}.
It could be closely related to the saddle-point singularity and the nesting of Fermi surface (FS) near van Hove filling \cite{YuSL,WangWS,Kiesel}. 
Interestingly, the formation of superlattice in $A$V$_{3}$Sb$_{5}$ below CDW transition temperature $T_{\rm CDW}\sim$ 80 K - 110 K \cite{Ortiz1,Ortiz2,Ortiz3,YinQW} results in the appearance of charge gap observed in optical spectroscopy but the softening of acoustic phonon is absent near the CDW vector \cite{Uykur,ZhouX,LiHX}. In addition, another 4$a_{0}$ unidirectional superlattice has also been observed at low temperature \cite{ChenH,ZhaoH}.
On the other hand, superconductivity in these materials also exhibits some intricate relationship with the CDW state and the nature of superconductivity in these materials is still under debate \cite{LiangZ,ChenH,MuC,ZhaoCC,DuanW}. For example, the $T_{\rm CDW}$ decreases with pressure when the superconducting transition temperature $T_{c}$ shows an unusual multiple-dome feature with a significant enhancement to about 8 K at 2 GPa \cite{ZhaoCC,ChenKY,ChenX,ZhangZ}.

In comparison to the intensive studies on the CDW state and superconductivity of $A$V$_{3}$Sb$_{5}$, the experimental investigations of topological features of these materials are still scarce. 
In this work, we present a detailed study on transport properties of CsV$_{3}$Sb$_{5}$ single crystals in a magnetic field up to 32 T.
The analysis of Shubnikov-de Haas (SdH) oscillations of $ab$-plane resistivity $\rho_{xx}$ and the results of theoretical calculations using the structure with in-plane ISD distortion reveal the existence of several extremal orbits of FSs with relatively small cyclotron masses $m^{*}$s and nonzero Berry phases, providing a direct evidence for the existence of nontrivial topological electronic structures in CsV$_{3}$Sb$_{5}$ at CDW state.	

\begin{figure}
\centerline{\includegraphics[scale=0.160]{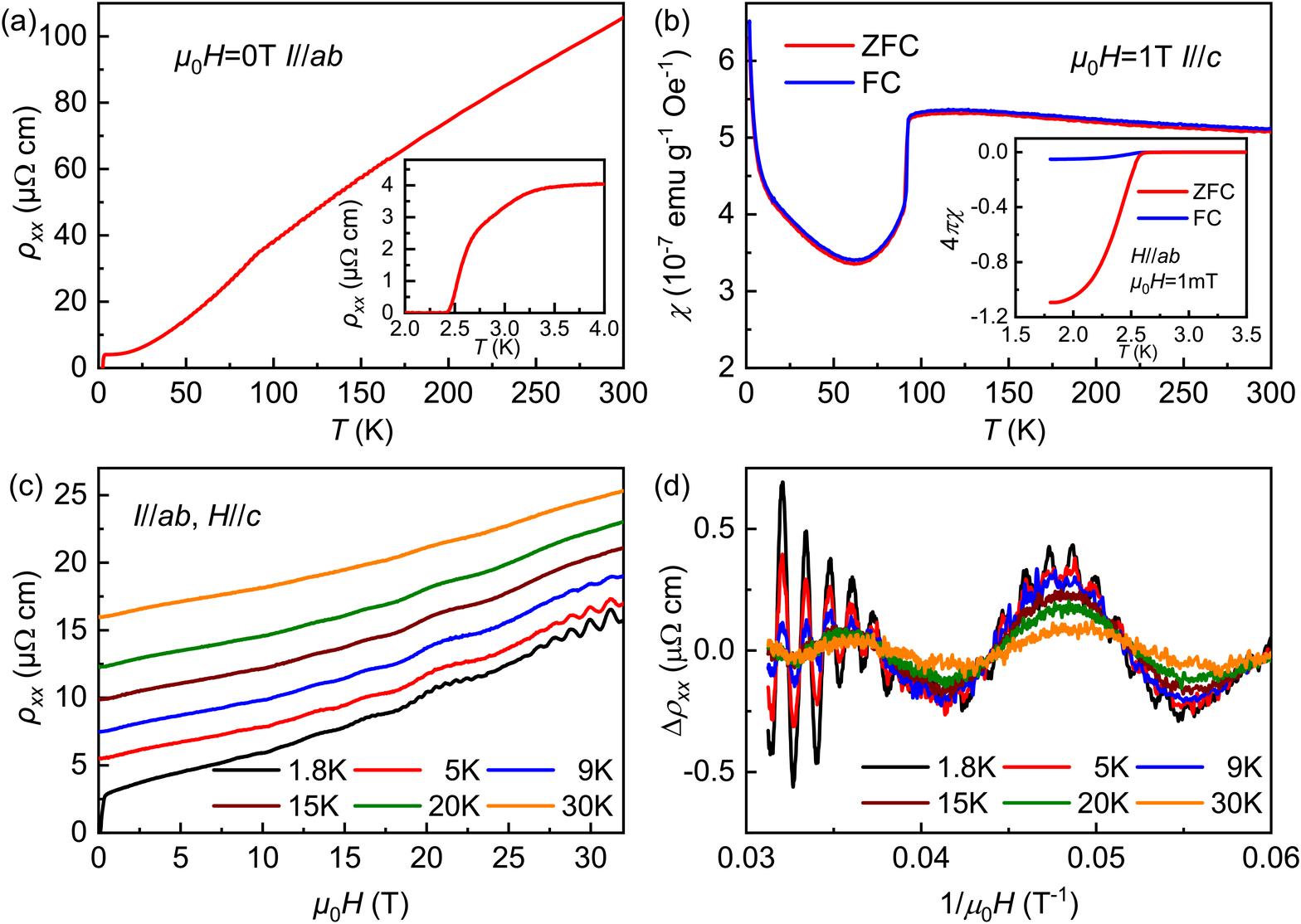}} \vspace*{-0.3cm}
\caption{(a) Temperature dependence of $\rho_{xx}(T)$. Inset: enlarged $\rho_{xx}(T)$ curve at low temperature. (b) Temperature dependence of $\chi(T)$ at $\mu_{0}H=$ 1 T for $H\Vert c$ with ZFC and FC modes. Inset: ZFC and FC $4\pi\chi(T)$ curves at $\mu_{0}H=$ 1 mT for $H\Vert ab$ below 3.5 K. (c) Field dependence of $\rho_{xx}(\mu_{0}H)$ for $H\Vert c$ at various temperatures. The $\rho_{xx}(\mu_{0}H)$ curves are shifted vertically for clarity. (d) The oscillatory component $\Delta\rho_{xx}$ as the function of 1/$\mu_{0}H$ at various temperatures. The $\Delta\rho_{xx}$ is extracted from $\rho_{xx}$ by subtracting a smooth background.}
\end{figure}

CsV$_{3}$Sb$_{5}$ single crystals were grown by the self-flux method. The detailed methods of experimental characterizations and theoretical calculations are shown in Supplemental Material (SM) \cite{SM}. 
The $\rho_{xx}(T)$ of CsV$_{3}$Sb$_{5}$ single crystal exhibits a metallic behavior with a kink at $T_{\rm CDW}\sim$ 92 K due to the CDW transition (Fig. 1(a)) \cite{Ortiz2}, which also leads to a sharp drop of magnetic susceptibility $\chi(T)$ at the same temperature (Fig. 1(b)).
The nearly overlapped zero-field-cooling (ZFC) and field-cooling (FC) $\chi(T)$ curves suggest that this anomaly should be ascribed to the charge ordering transition.
At low temperature, superconductivity appears at $T_{c}\sim$ 2.5 K, leading to the zero-resistance and diamagnetic transitions (insets of Figs. 1(a) and 1(b)), similar to previous result \cite{Ortiz2}.
At $T=$ 1.8 K, the superconducting volume fraction (SVF) estimated from the ZFC $4\pi$$\chi(T)$ curve for $H\Vert ab$ is about 109 \%, indicating a bulk superconductivity of CsV$_{3}$Sb$_{5}$ single crystal. 
The small SVF obtained from the FC $4\pi\chi(T)$ curve implies the relatively strong flux pinning effect in CsV$_{3}$Sb$_{5}$. 
Figure 1(c) shows the field dependence of $\rho_{xx}(\mu_{0}H)$ up to 32 T for $H\Vert c$ at various temperatures. At 1.8 K, there is a fast increase of $\rho_{xx}(\mu_{0}H)$ at low-field region due to the suppression of superconductivity under field. 
At higher fields, all of $\rho_{xx}(\mu_{0}H)$ curves exhibit a positive magnetoresistance without saturation up to 32 T. The most prominent feature of $\rho_{xx}(\mu_{0}H)$ curves is the appearance of SdH oscillations at high-field region, which decays with increasing temperature.
After subtracting a smooth background, the oscillatory parts of $\Delta \rho_{xx} (= \rho_{xx} - \langle \rho_{xx}\rangle$) as a function of $1/\mu_{0}H$ exhibit complex periodic behaviors (Fig. 1(d)), indicating the contributions of several frequency components. The low-frequency SdH oscillations are still observable at 30 K, while the high-frequency ones disappear at 15 K.

\begin{figure}
\centerline{\includegraphics[scale=0.160]{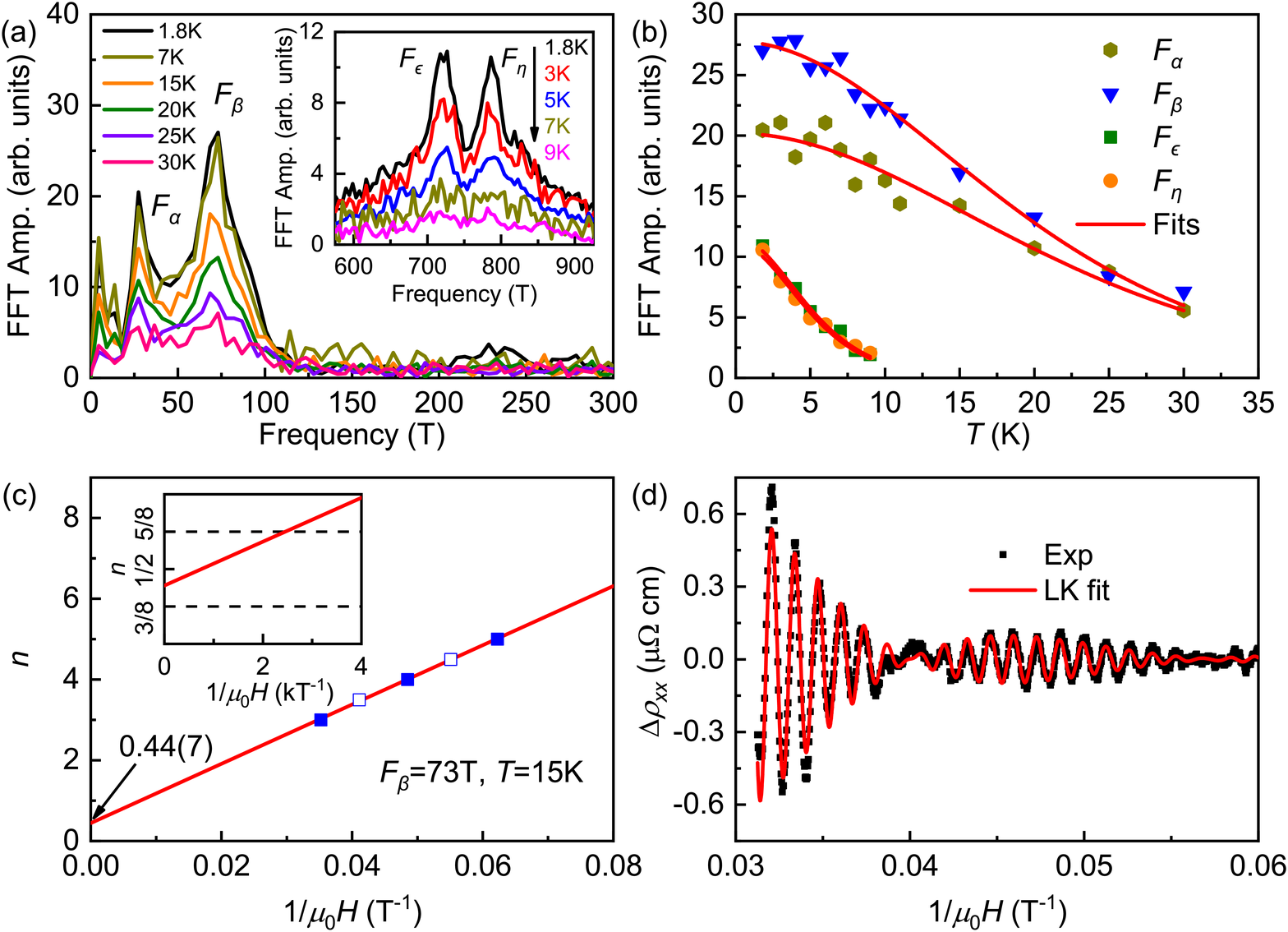}} \vspace*{-0.3cm}
\caption{(a) FFT amplitudes of the SdH oscillations at low-frequency region. Inset: FFT amplitudes of the SdH oscillations at high-frequency region. (b) Temperature dependence of FFT amplitudes of $F_{\alpha}$, $F_{\beta}$, $F_{\epsilon}$, and $F_{\eta}$ peaks. The solid lines represent the L-K formula fits for $m^{*}$s. (c) Landau index $n$ vs. 1/$\mu_{0}H$ for the frequency of $F_{\beta}$, derived from the $\Delta\rho_{xx}$ curve at 15 K. The closed squares denote the integer index ($\Delta\rho_{xx}$ peak), and the open squares indicate the half integer index ($\Delta\rho_{xx}$ valley). Inset: magnified view around the intercept. (d) The high-frequency oscillatory components at 1.8 K. The red line represents the fit using the two-frequency L-K formula.
}
\end{figure}

As shown in Fig. 2(a) and its inset, the fast Fourier transform (FFT) spectra of SdH oscillations between 4 T and 32 T for $H\Vert c$ reveal several frequencies at $F_{\alpha}=$ 27 T, $F_{\beta}=$ 73 T, $F_{\epsilon}=$ 727 T and $F_{\eta}=$ 786 T. The two low-frequency peaks are close to previously reported results \cite{YuFH}.
The peak at 5 T is excluded for the further analysis because of its limited number of periods in the field range from 4 T to 32 T.
According to the Onsager relation $F=(\hbar/2\pi e)A_{F}$ where $A_{F}$ is the area of extremal orbit of FS \cite{Shoenberg}, the determined $A_{F}$s are 0.00258, 0.00697, 0.06940 and 0.07503 \AA$^{-2}$ for $F_{\alpha}$, $F_{\beta}$, $F_{\epsilon}$ and $F_{\eta}$, respectively.
The $A_{F_{\alpha}}$ and $A_{F_{\beta}}$ are very small, taking only about 0.17 \% and 0.46 \% of the whole area of BZ in the $k_{x}$-$k_{y}$ plane when taking $a=$ 5.50552 \AA\ (Fig. S1 of SM) \cite{SM}. In contrast, $A_{F_{\epsilon}}$ and $A_{F_{\eta}}$ take about 4.61 \% and 4.99 \% of the Brillouin zone (BZ) area.

In general, the SdH oscillations with several frequencies can be described by linear superposition of the multifrequency Lifshitz-Kosevich (L-K) formula, and each of which can be expressed as \cite{Shoenberg,HuJ},
\begin{equation}
\Delta \rho_{xx}^{i} \propto \frac{5}{2} \sqrt{\frac{\mu_{0}H}{2 F}} R_{T} R_{D} R_{S} \cos [2 \pi(F/\mu_{0}H+\gamma-\delta+\varphi)],
\end{equation}
where for the $i$th $\mathrm{SdH}$ oscillation component, $F$ is frequency, $R_{T}=(\lambda m^{*}T/\mu_{0}H)/\sinh(\lambda m^{*}T/\mu_{0}H)$, $\quad R_{D}=\exp(-\lambda m^{*} T_{D}/\mu_{0}H)$, $\quad R_{S}=\cos(\pi m^{*} g^{*})$, $m^{*}$ is cyclotron mass in unit of free electron mass $m_{e}$, $T_{D}$ is the Dingle temperature, $g^{*}$ is the effective $g$ factor, and constant $\lambda=2\pi^{2}k_{B}m_{e}/e\hbar\approx$ 14.7 T/K. The phase factor $\gamma-\delta+\varphi$ contains $\gamma=1/2-\phi_{B}/2\pi$ where $\phi_{B}$ is Berry phase, $\delta$ is determined by the dimensionality of FS ($\delta=0$ and $\pm1/8$ for the respective two-dimensional (2D) and 3D cases) \cite{Shoenberg,Mikitik,Lukyanchuk}, and $\varphi=1/2$ ($\rho_{xx}\gg\rho_{yx})$ or 0 ($\rho_{xx}\ll\rho_{yx})$ \cite{XiangFX}.
With the $\mu_{0}H_{\rm avg}$ (= 18 T) being the average value of the field window used for the FFT of SdH oscillations \cite{Rhodes}, the fitted $m^{*}$s from the temperature dependence of the FFT amplitudes to the $R_{T}$ (Fig. 2(b)) are 0.127(8), 0.142(4), 0.54(3), and 0.55(3) $m_{e}$ for $F_{\alpha}$, $F_{\beta}$, $F_{\epsilon}$, and $F_{\eta}$, respectively. The $m^{*}$s for $F_{\alpha}$ and $F_{\beta}$ are close to those in KV$_{3}$Sb$_{5}$ and RbV$_{3}$Sb$_{5}$ \cite{YangSY,YinQW}, but larger than previously reported values for CsV$_{3}$Sb$_{5}$ \cite{YuFH}.
The Landau level (LL) fan diagram for the frequency $F_{\beta}$ is constructed from the SdH oscillations at 15 K (Fig. 2(c)) because the high-frequency oscillations have damped completely at this temperature. 
For $\rho_{xx}/|\rho_{yx}|\sim$ 5 for $\mu_{0}H<$ 14 T (Fig. S2 of SM) \cite{SM} and $\Delta\rho_{xx}/\rho_{xx}<$ 0.04, the longitudinal conductivity $\sigma_{xx} (=\frac{\rho_{xx}}{\rho_{xx}^{2}+\rho_{yx}^{2}}) \approx \frac{1}{\rho_{xx}}$, meaning that the minima and maxima of the oscillations in the $\sigma_{xx}$ are out of phase with those in the $\rho_{xx}$. Correspondingly, the LL integer $n$ should be assigned to the oscillatory maxima of $\rho_{xx}$ while the LL half-integer index $n+1/2$ is assigned to the oscillatory minimum of $\rho_{xx}$ \cite{XiangFX}. 
From the linear fit of the $n$ as a function of 1/$\mu_{0}H$, the intercept on the LL index axis is 0.44(7) in between 3/8 and 5/8, indicating a nontrivial $\phi_{\rm B}$ [= (1 - 0.44(7))$\times$2$\pi$ = 1.1(1)$\pi$] for $F_{\beta}$ as expected in a Dirac system \cite{Mikitik}. This value also implies that the $\delta$ deviates from the 3D limit $|\delta|=$ 1/8, probably because of the quasi-2D nature of corresponding FS shown below. 
Moreover, the fitted oscillation frequency is 73(1) T, nearly same as the value shown in Fig. 2(b). It suggests the reliability of linear fit in the LL fan diagram.
In addition, the Berry phases should also be obtained from multi-frequency L-K formula fit. When taking the values of frequencies and $m^{*}$s obtained from the FFT spectrum for $F_{\alpha}$ and $F_{\beta}$, the SdH oscillation at 15 K can be fitted very well by using the two-frequency L-K formula (Fig. S3 of SM) \cite{SM}. The fitted $\phi_{\rm B}$ of $F_{\beta}$ is 1.076(4)$\pi$, perfectly consistent with above value obtained from the LL fan diagram, and for $F_{\alpha}$ the value of $\phi_{\rm B}$ is 0.720(8)$\pi$.
On the other hand, because two frequencies of $F_{\epsilon}$ and $F_{\eta}$ form a beat frequency, it is hard to analyse their Berry phases using the LL fan diagram.
Instead, the Berry phases are obtained from the two-frequency L-K formula fit. In order to achieve a more accurate fit, we have separated the high-frequency SdH oscillations by band pass filter (200 - 1200 T). When using the values of frequencies and $m^{*}$s for $F_{\epsilon}$ and $F_{\eta}$ obtained from above analysis and setting $T=$ 1.8 K, the high-frequency components of SdH can be fitted very well (Fig. 2(d)).
The fit gives $(\phi_{\rm B}/2\pi+\delta)_{\epsilon}$ = 0.332(3) and $(\phi_{\rm B}/2\pi+\delta)_{\eta}$ = 0.194(4). It implies that the $\phi_{\rm B}$ for $F_{\epsilon}$ is close to $\pi$ with $\delta\sim$ -1/8, i.e., the FS related to $F_{\epsilon}$ could have a nontrivial Berry phase and a 3D shape. For $F_{\eta}$, the $\phi_{\rm B}$ is close to 0 with $\delta\sim$ 1/8, meaning that the corresponding FS may be trivial.
Thus, the SdH oscillations for $F_{\beta}$ and $F_{\epsilon}$ with relatively small $m^{*}$s and nontrivial $\phi_{B}$s clearly indicate the existence of Dirac cones in CsV$_{3}$Sb$_{5}$.

\begin{figure}
\centerline{\includegraphics[scale=0.15]{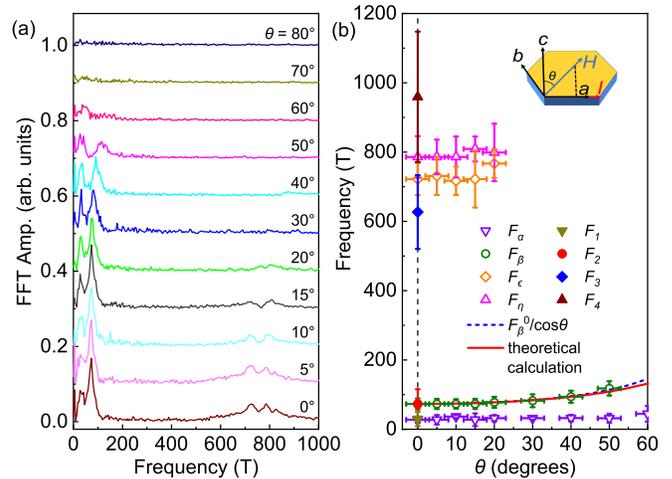}} \vspace*{-0.3cm}
\caption{(a) Field dependence of FFT amplitudes of SdH oscillations at various field directions ($\theta$s) when $T=$ 1.8 K. The data at different field directions have been shifted for clarity. (b) Angular dependence of oscillation frequencies $F_{\alpha}$, $F_{\beta}$, $F_{\epsilon}$, and $F_{\eta}$. Inset in (b) shows the definition of $\theta$. Error bars are defined as the half width at the half-height of FFT peaks and the angle uncertainty of measurement ($\sim$ 3$^{\circ}$). The blue dashed line is calculated from the formula $F_{\beta}(\theta)=F_{\beta}(0^{\circ})/\cos\theta$. The solid symbols at $\theta=$ 0 represent the frequencies calculated from the cross sections of unfolded FSs in pristine BZ shown in Figs. 4(b) and 4(c). The error bars originate from the uncertainties when evaluating the areas of cross sections of unfolded FSs. 
The red solid line is obtained from theoretical calculations using the FS in the reconstructed BZ.
}
\end{figure}

The anisotropic characteristics of the FSs of CsV$_{3}$Sb$_{5}$ are further clarified by investigating the relationship between $\Delta\rho_{xx}$ and 1/$\mu_{0}H$ as well as corresponding SdH oscillation frequencies at 1.8 K with various field directions (Fig. S4 of SM and Fig. 3(a)) \cite{SM}. The measurement configuration is shown in the inset of Fig. 3(b).
Because there is a peak broadening or splitting for $\theta=$ 40$^{\circ}$, 50$^{\circ}$ and 60$^{\circ}$, the Gaussian fits of peaks are used to determine the values of $F_{\alpha}$ and $F_{\beta}$ at those field directions. As shown in Fig. 3(b), The $F_{\alpha}$ depends on the field direction weakly when $\theta\leq$ 50$^{\circ}$ but there is an obvious shift of $F_{\alpha}$ to higher frequency at $\theta=$ 50$^{\circ}$. 
For the $F_{\beta}$, it moves to higher frequency with increasing $\theta$ and the peak is barely observed when $\theta>$ 50$^{\circ}$.
The angular dependence of the $F_{\beta}$ can be well fitted up to $\theta=$ 50$^{\circ}$ by using the formula $F_{\beta}(\theta)=F_{\beta}(0^{\circ})/\cos\theta$, where $F_{\beta}(0^{\circ})$ is the position of $F_{\beta}$ for $H\Vert c$ (blue dashed line in Fig. 3(b)). This result implies that the FS corresponding to $F_{\beta}$ has a quasi-2D or prolate ellipsoid shape.
On the other hand, for $F_{\epsilon}$ and $F_{\eta}$, the peak positions also shift to higher frequency slightly, but the peak amplitudes decay much more quickly than those of $F_{\alpha}$ and $F_{\beta}$. The former ones become hard to distinguish when $\theta>$ 20$^{\circ}$.

\begin{figure}
\centerline{\includegraphics[scale=0.37]{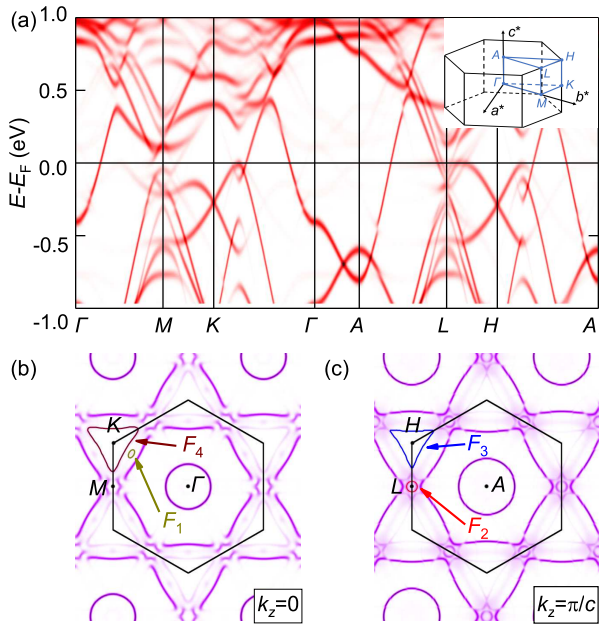}} \vspace*{-0.3cm}
\caption{(a) Calculated unfolded band structure of CsV$_{3}$Sb$_{5}$ along high-symmetry directions across the pristine BZ using the structure with an ISD distortion. 
Inset: schematic of the 3D BZ in the pristine phase. High-symmetry points and momentum lines are marked. 
(b) and (c) The cross sections of unfolded FSs at $k_{z}=$ 0 and $\pi/c$ planes in the pristine BZ, respectively. 
The extremal orbits related to $F_{1}$ - $F_{4}$ are highlighted.
}
\end{figure}

In order to make a comprehensive understanding on experimental results of SdH oscillations, the first-principles electronic structure calculations were carried out based on the structure at CDW state in which there is an ISD distortion in kagome lattice \cite{JiangYX,TanH,LiangZ,LiHX}. 
Due to the weak interlayer interaction, only the in-plane distortion, i.e., 2$\times$2$\times$1 superlattice, is considered during theoretical calculation for simplicity of analysis.
The unfolded band structure in CDW state with the ISD distortion along the high-symmetry paths in the pristine BZ is shown in Fig. 4(a) and it is similar to the previous report \cite{TanH}. 
It is noted that the $E_{\rm F}$ is shifted down slightly by 33.3 meV in order to match the experimental results of SdH oscillation. 
When compared with the pristine phase without structural distortion (Fig. S5 of SM) \cite{SM}, the general feature of band structure at CDW state is unchanged, especially for those bands originating from the Sb orbitals such as the band near $\Gamma$ point.
Moreover, the Dirac cones at $K$ and $H$ points of BZ below $E_{\rm F}$ are also almost intact, which have been confirmed by the APRES measurements \cite{LouR}.
However, the bands near the BZ boundary such as $M$ and $L$ points mainly contributed from the V orbitals are modified obviously, which is reasonable because the 2$\times$2 ISD distortion appears in the V-kagome lattice.

Figures 4(b) and 4(c) show the cross sections of unfolded FSs at $k_{z}=$ 0 and $\pi/c$ planes in the pristine BZ. 
There are a tiny oval cross section of FSs (labelled as $F_{1}$) in the $\Gamma-K$ line and a small circular-shaped one (labelled as $F_{2}$) appearing around the $L$ point. 
In contrast, two FSs at $K$ and $H$ points show a rounded triangular shape (labelled as $F_{4}$ and $F_{3}$). 
On the other hand, the cross sections of FSs locating at the center of BZ ($\Gamma$ and $A$ points) have circular shapes with different radii. 
Moreover, there is a large hexagonal-shaped cross section of FS around the $A$ point and it becomes a discontinued hexagon when moving to the $k_{z}=$ 0 plane. 
Above results are consistent with the angle-resolved photoemission spectroscopy (ARPES) results in principle \cite{Ortiz2,Nakayama,LouR}. 
Assuming the extremal orbits contributed from these cross sections of FSs, the corresponding frequencies below 1000 T are calculated and shown in Fig. 3(b). 
It is found that the experimental $F_{\alpha}$ - $F_{\eta}$ could be assigned to the calculated FS sections of $F_{1}$ - $F_{4}$ (Figs. 4(b) and 4(c)).
Taking $F_{\beta}$ as an example, the validity of assignment is further confirmed by calculating angular dependence of frequencies using the FS in the reconstructed BZ (Fig. S6 of SM) \cite{SM}. 
As shown in Fig. 3(b), the theoretical curve (red solid line) can describe the trend of experimental result very well. Because the FS related to $F_{\beta}$ is a prolate ellipsoid in shape centering at $L$ point (see the red pocket in Fig. S6 of SM) \cite{SM}, this explains the similar trend of theoretical curve and that calculated using the formula $F_{\beta}(\theta)=F_{\beta}(0^{\circ})/\cos\theta$ when $\theta\leq$ 50$^{\circ}$. In addition, the calculated $m^{*}$ is 0.145 $m_{e}$, which is also in good agreement with the experimental value of $F_{\beta}$ (0.142(4) $m_{e}$).
Such small $m^{*}$ reflects the linear dispersion of this electronic pocket at $L$ point. 
Importantly, according to previous theoretical calculations \cite{TanH}, The ISD phase in the CDW state has a topologically nontrivial band structure with nonzero $Z_{2}$ topological invariants for the bands near $E_{\rm F}$. Thus, it will lead to a $\pi$ Berry phase accumulated along the cyclotron orbit, consistent with present experimental value of $\phi_{\rm B}$.
On the other hand, the differences between $F_{\epsilon}$ ($F_{\eta}$) and the calculated value for $F_{3}$ ($F_{4}$) may be due to the influence of lattice distortion along the $c$ direction at CDW state with 3D 2$\times$2$\times$2 superlattice \cite{LiangZ,LiHX}. In that case, the band folding also happens along the $k_{z}$ direction and the $F_{3}$ and $F_{4}$ will overlap each other, possibly leading to the modifications of both cross sections of FSs, i.e., the changes of corresponding frequencies. Further study is needed in order to clarify this issue.

It is worth mentioning that another study \cite{Ortiz4} on the SdH oscillation of CsV$_{3}$Sb$_{5}$ with the field up to 14 T reported in very recent shows similar results about the Fermiology of CsV$_{3}$Sb$_{5}$ to present work. Although the calculated frequencies of extremal orbits are different slightly, the qualitative consistence between experiments and theoretical calculations in these two studies suggests that the structure with an ISD distortion could be the most stable structure at CDW state.

In summary, the high-field transport measurements at low temperature indicate that there are several SdH oscillation frequencies with relatively small $m^{*}$s for CsV$_{3}$Sb$_{5}$. Theoretical calculations using the structure with an ISD distortion at CDW state suggest that the frequencies of SdH oscillations could be related to the extremal orbits of FSs near the high-symmetry points at BZ boundary. Importantly, the $\phi_{B}$s of FSs related to the frequencies of $F_{\beta}$ and $F_{\epsilon}$ are close to $\pi$, confirming the predicted nontrivial topological properties of CsV$_{3}$Sb$_{5}$ at CDW state.
Thus, the kagome metals $A$V$_{3}$Sb$_{5}$ pave a new way to study the correlation effects on topological electronic structures.

This work was supported by Beijing Natural Science Foundation (Grant No. Z200005), National Key R\&D Program of China (Grant No. 2018YFE0202600, 2017YFA0302903 and 2016YFA0300504), National Natural Science Foundation of China (Grant No. 11822412, 11774423, U1932217 and U2032215), the Fundamental Research Funds for the Central Universities and Research Funds of Renmin University of China (RUC) (Grant No. 18XNLG14, 19XNLG13 and 19XNLG17), Beijing National Laboratory for Condensed Matter Physics, and Youth Innovation Promotion Association of CAS (Grant No. 2017483). Computational resources were provided by the Physical Laboratory of High Performance Computing at Renmin University of China and Shanghai Supercomputer Center.

$^{\dag}$ Y.F., N.N.Z, Z.C. and Q.W.Y contributed equally to this work.

$\ast$ Corresponding authors: xdzhu@hmfl.ac.cn (X. D. Zhu); kliu@ruc.edu.cn (K. Liu); hlei@ruc.edu.cn (H. C. Lei).

\end{document}